\documentclass[10pt,twocolumn,letterpaper]{article}

\usepackage{cvpr}
\usepackage{times}
\usepackage{epsfig}
\usepackage{graphicx}
\usepackage{amsmath}
\usepackage{amssymb}

\usepackage{array}
\usepackage{stfloats}
\usepackage{pifont}
\newcommand{\xmark}{\ding{55}}%
\usepackage[labelformat=empty]{subfig}

\usepackage[breaklinks=true,bookmarks=false]{hyperref}

\cvprfinalcopy 

\def\cvprPaperID{70} 

\ifcvprfinal\fi
\begin{document}

\title{Perceptual Extreme Super Resolution Network with Receptive Field Block}


\author{
\hspace{0mm}Taizhang Shang
\hspace{10mm}Qiuju Dai
\hspace{10mm}Shengchen Zhu
\hspace{10mm}Tong Yang\thanks{Corresponding author: yangtong@oppo.com}
\hspace{10mm}Yandong Guo
\vspace{1mm}
\\
OPPO Research, China
\\
\hspace{0mm}{\tt\small \{shangtaizhang, daiqiuju, zhushengchen, yangtong, guoyandong\}@oppo.com}
}

\maketitle

\begin{abstract}
Perceptual Extreme Super-Resolution for single image is extremely difficult, because the texture details of different images vary greatly. To tackle this difficulty, we develop a super resolution network with receptive field block based on Enhanced SRGAN. We call our network RFB-ESRGAN. The key contributions are listed as follows. First, for the purpose of extracting multi-scale information and enhance the feature discriminability, we applied receptive field block (RFB) to super resolution. RFB has achieved competitive results in object detection and classification. Second, instead of using large convolution kernels in multi-scale receptive field block, several small kernels are used in RFB, which makes us be able to extract detailed features and reduce the computation complexity. Third, we alternately use different upsampling methods in the upsampling stage to reduce the high computation complexity and still remain satisfactory performance. Fourth, we use the ensemble of 10 models of different iteration to improve the robustness of model and reduce the noise introduced by each individual model. Our experimental results show the superior performance of RFB-ESRGAN. According to the preliminary results of NTIRE 2020 Perceptual Extreme Super-Resolution Challenge, our solution ranks first among all the participants.
\end{abstract}

\section{Introduction}
Single image super-resolution (SISR) is a task to generate high-resolution (HR) image with a single low-resolution image. The algorithms of SISR can be divided into three categories: interpolation-based methods, reconstruction-based methods, and learning-based methods \cite{yang2019deep}. Interpolation-based SISR methods are very speedy and straightforward, such as bicubic interpolation \cite{keys1981cubic} and Lanczos
resampling \cite{duchon1979lanczos}. But some works have shown that interpolation
methods would lose the details of images \cite{yang2010image,dong2015image}.
Reconstruction-based SR methods \cite{dai2009softcuts,sun2008image,yan2015single} adopt sophisticated prior knowledge to restrict the possible solution space with the advantage of generating flexible and sharp details \cite{yang2019deep}. However, as the scale factor increases, the performance of reconstruction-based SR methods decreases, and reconstruction-based SISR methods typically cost a lot of time. Learning-based SISR methods usually use machine learning algorithms to get the model which produces the mapping from low resolution to high resolution images. The learning-based methods has attracted much attention owning to their outstanding performance and fast computation. Such as Markov random field method \cite{freeman2002example}, neighbor embedding method \cite{chang2004super}, sparse coding methods \cite{yang2010image,zeyde2010single,timofte2013anchored}, and random forest method \cite{schulter2015fast}. Recently, many deep learning based methods have been proposed to solve the SISR problem, and deep learning based SISR methods have demonstrated great superiority to other SISR methods.

Recently, deep learning algorithms have been widely used in different fields.  Super-resolution CNN (SRCNN) \cite{dong2014learning} is the first work to solve SISR problem using neural network method, it reportedly demonstrated vast superiority over traditional methods. The main reason it achieves good results is the CNN's strong capability of learning rich features from big data in an end-to-end manner. After SRCNN was proposed, VDSR \cite{kim2016accurate} further use deep model to solve SISR problem, it has 20 layers in the network. EDSR \cite{lim2017enhanced} proposed to remove the batch normalization (BN) layer in model, for BN layer will introduce a shift to the feature, and this shift may be harmful to the final performance. RCAN \cite{zhang2018image} was proposed using the channel attention in SISR problem. However, these methods' objective function has largely focused on minimizing the mean squared reconstruction error, which lead to the SR results lack of high-frequency details. To address this problem, super-resolution using generative adversarial network (SRGAN) \cite{ledig2017photo} has been proposed, which can recover the finer texture details even with large upscaling factors. Enhanced super-resolution generative adversarial networks (ESRGAN) \cite{wang2018esrgan} was proposed to further improve the performance of deep learning based SISR model. With the powerful feature extraction capabilities of deep learning models and the generative adversarial method, deep learning-based methods can effectively recover the finer details and textures.

NTIRE 2020 Perceptual Extreme Super-Resolution Challenge, that is, the task of super-resolving (increasing the resolution) an input image with a magnification factor $\times$16 based on a set of prior examples of low and corresponding high resolution images.The aim is to obtain a model capable to produce high resolution results with the best perceptual quality and similarity to the ground truth. There are two difficulties in this challenge. First, we need to develop a model that can effectively recover the finer details and textures of low resolution image, and make the results be both photo-realistic and with high perceptual quality. Second, we need to minimize time complexity as much as possible while keep the satisfactory results at the same time.

In this work, we proposed to use multi-scale Receptive Fields Block (RFB) in the generative network to restore the finer details and textures of the super-resolution image. RFB can extract different scale features from previous feature map, which means it can extract the coarse and fine features from input LR images. To reduce time complexity and still maintain satisfactory performance, RFB use several small kernels instead of large kernels, and we alternately use different upsampling methods in um-sampling stage of the generative network. Finally in the testing phase, we use model fusion to improve the robustness and stability of the model to different test images.


\section{Related Work}
\textbf{Single Image Super-Resolution}. Since the pioneer work of SRCNN \cite{dong2015image}, deep learning based methods have brought significant improvement in image super-resolution \cite{kim2016accurate, lim2017enhanced, zhang2018image, ledig2017photo, wang2018esrgan}. For image super-resolution, VDSR \cite{kim2016accurate} reveals that increasing network depth shows a significant improvement in SISR. EDSR \cite{lim2017enhanced} abandoned batch normalization (BN) layers to prevent BN artifacts of SR images. Perceptual loss \cite{gatys2015neural} was first proposed in the field of style transfer. SRGAN \cite{ledig2017photo} use the perceptual loss to reduce the gap between SR images and human visual perception, and achieved very good results. ESRGAN \cite{wang2018esrgan} introduced the Residual-in-Residual Dense Block (RRDB) into generative network, and proposed to let the discriminator predict relative realness instead of the absolute value in SISR. Our RFB-ESRGAN use a deep neural network without BN layers as the backbone of the generative network, also benefit from the RRDB and use relative realness in loss function instead of the absolute value.

\textbf{Multi-scale Receptive Fields}. GoogleNet \cite{szegedy2015going} increase the width of the network in classification field, use multi-scale kernels to extract different scale features. After the pioneer work of GoogleNet, many other deep networks have tried to use multi-scale kernels to increase the diversity of features of the network in different network structures, and achieved good results. Inspired by the multi-scale kernels and the structure of Receptive Fields (RFs) in human visual systems, RFB-SSD \cite{liu2018receptive} proposed Receptive Fields Block (RFB) for object detection. In our work,  we introduce RFB into our generative network for super-resolution. 

\textbf{Upsampling Methods}. In the early deep learning based SISR, most works put the upsampling stage in the front of the models, like SRCNN \cite{dong2015image}, VDSR \cite{kim2016accurate}. It will make the model very large, and cost a lot of time in test phase. FSRCNN \cite{dong2016accelerating} make the upsampling stage in the end of the model,  this make the input size more small and the model more deeper possible. FSRCNN use deconvolution for upsampling, while ESRGAN \cite{wang2018esrgan} and some other works use nearest interpolation for upsampling. ESPCN \cite{shi2016real} proposed the sub-pixel method for upsampling to reduce the time complexity. For RFB-ESRGAN, we alternately use nearest interpolation and sub-pixel convolution for upsampling. Here is our thought, nearest interpolation method focus on the computation in space dimension, while the sub-pixel convolution method focus on the computation in depth dimension. The alternative use of them allows for full communication of information between depth and space.

\textbf{Minimize Time Complexity}. For the purpose of minimize time complexity many networks design tricks have been proposed. GoogleNet \cite{szegedy2015going} uses bottleneck layers to reduce the time complexity. MobileNet \cite{howard2017mobilenets} uses depth-wise separable convolution to speed up the model running on edge devices. In our work, RFB uses small kernels to instead of large kernels, and we also alternately use nearest and sub-pixel methods in upsampling stage. Thus, we can minimize the time complexity of the model as much as possible while keep satisfactory performance at the same time.

\section{Super Resolution Network with Receptive Field Block}
Extreme single image super-resolution reconstruction aims to recover lost high-frequency (rich detail) while maintaining content consistency \cite{gu2019aim}. Most SR network architectures are designed based on improving the PSNR (Peak Signal-to-Noise Ratio) value. However, the images reconstructed by PSNR-oriented methods are particularly smooth and lack high-frequency details. Perceptual-driven methods have been proposed to improve perceptual quality of SR results. Generative adversarial network \cite{goodfellow2014generative} is introduced to SR to generate results more naturally. SRGAN \cite{ledig2017photo} and ESRGAN \cite{wang2018esrgan} significantly improves the overall perceptual quality of SR outputs over PSNR-oriented methods. We proposed a novel Super Resolution Network based on ESRGAN named RFB-ESRGAN.

\subsection{Basic Network Architecture}
\begin{figure*}[htbp]
\centering 
\includegraphics[width=0.9\textwidth]{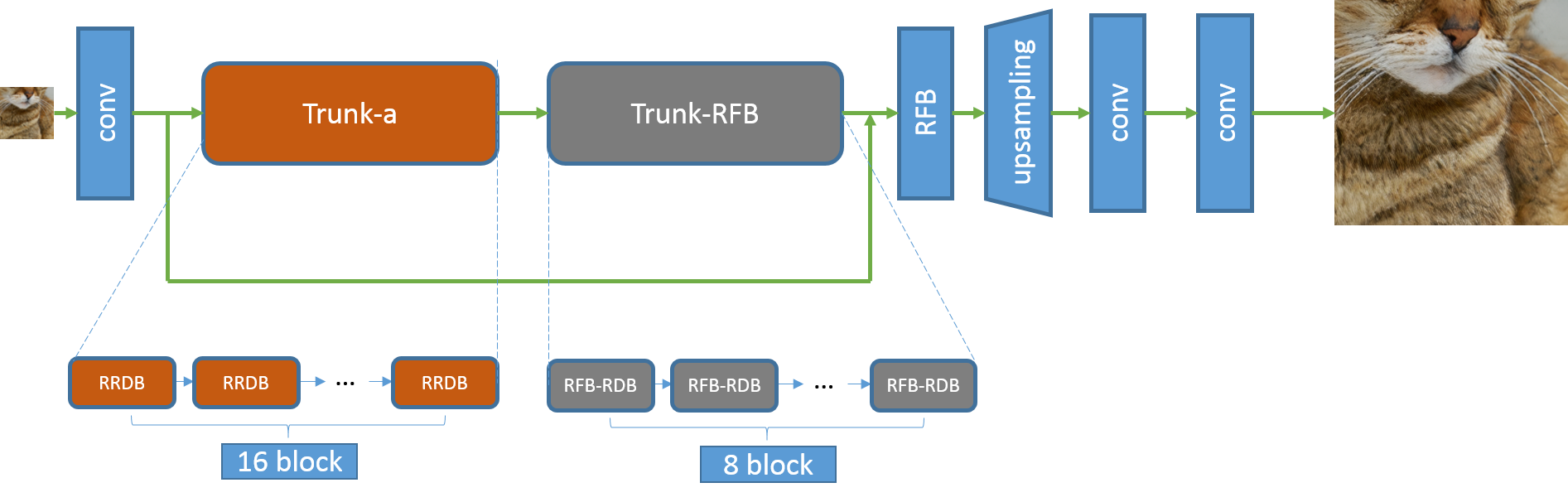}
\caption{The structure of RFB-ESRGAN.}
\label{fig:RFB-ESRGAN}
\end{figure*}

The proposed network structure consists of 5 parts shown in Fig. \ref{fig:RFB-ESRGAN}, namely the first convolution module, the Trunk-a module, the Trunk-RFB module, upsampling module and the final convolution module.

\begin{figure}[htbp]
\centering 
\includegraphics[width = 0.5\textwidth]{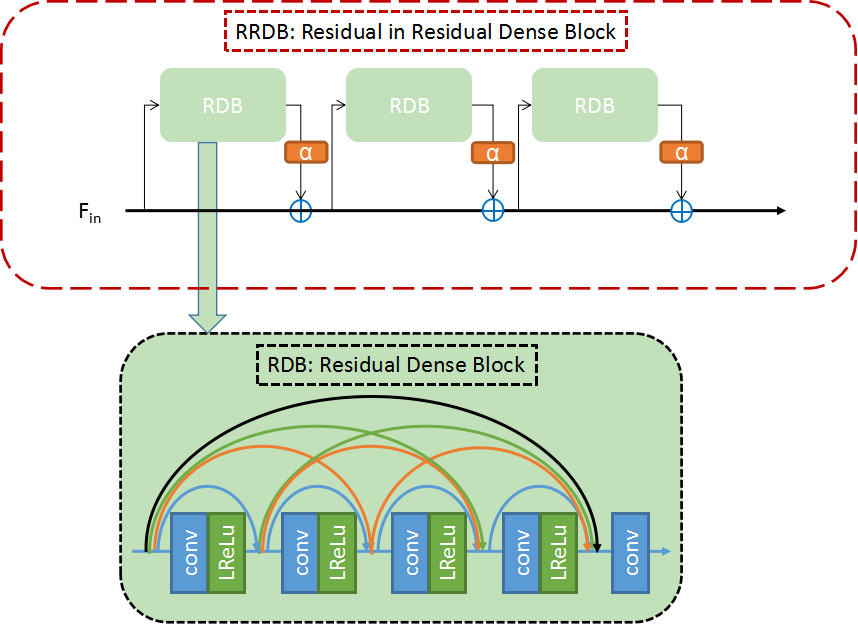}
\caption{Residual in Residual Dense Block (RRDB).}
\label{fig:RRDB}
\end{figure}

\begin{figure}[htbp]
\centering 
\includegraphics[width = 0.5\textwidth]{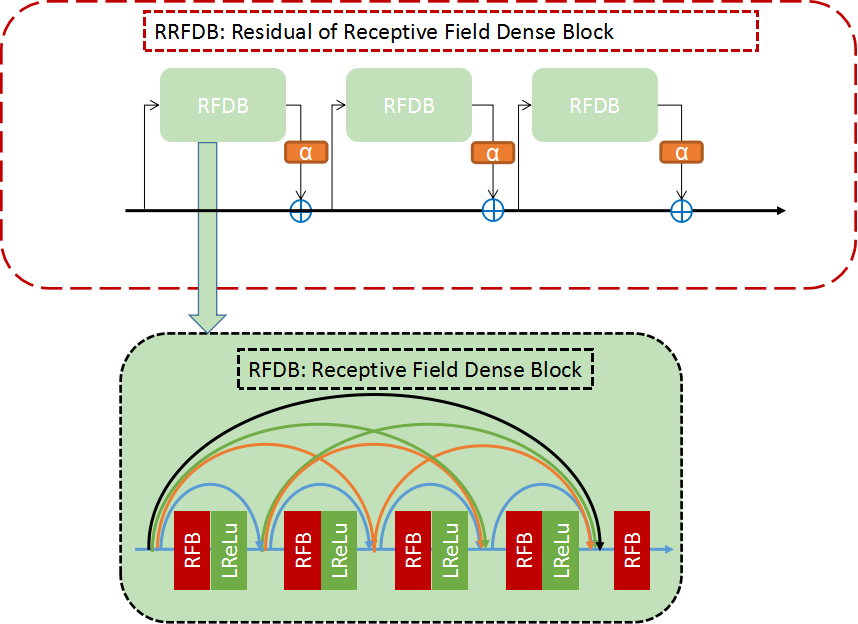}
\caption{Residual of Receptive Field Dense Block (RRFDB).}
\label{fig:RRFDB}
\end{figure}

The first convolution module is a convolution layer with a kernel size of $3\times3$, which can be formulated as equation \eqref{eq:conv}. where $f_{conv}$ denotes the first convolution function for the input LR image $I_{LR}$.
\begin{equation}
    x_{conv}=f_{conv}(I_{LR}) \label{eq:conv}
\end{equation}

 Trunk-a module consists of 16 RRDBs (Fig. \ref{fig:RRDB}). Define the function of nth RRDB in Trunk-a as $f_{RRDB}^n$. Trunk-a output can be given by the follow formula \eqref{eq:RRDB}.
\begin{equation}
    x_{RRDB}=f_{RRDB}^n(f_{RRDB}^{n-1}(...f_{RRDB}^0(x_{conv})...) \label{eq:RRDB}
\end{equation}
For perceptual extreme SR task we introduced RRFDBs (Fig. \ref{fig:RRFDB}) in our work, where we assemble RFB\cite{liu2018receptive} (Fig. \ref{fig:RFB}) in it. The RFB is consist of vary scale convolution filters, such we can restore rich image details for super resolution. Define the function of $m$th RRFDB in Trunk-RFB as $f_{RRFDB}^m$. The output of several stacked RRFDBs can be given by equation \eqref{eq:RRFDB}.
\begin{equation}
    x_{RRFDB}=f_{RRFDB}^m(f_{RRFDB}^{m-1}(...f_{RRFDB}^0(x_{RRDB})...) \label{eq:RRFDB}
\end{equation}

\begin{figure}[htbp]
\centering 
\includegraphics[height=0.4\textwidth]{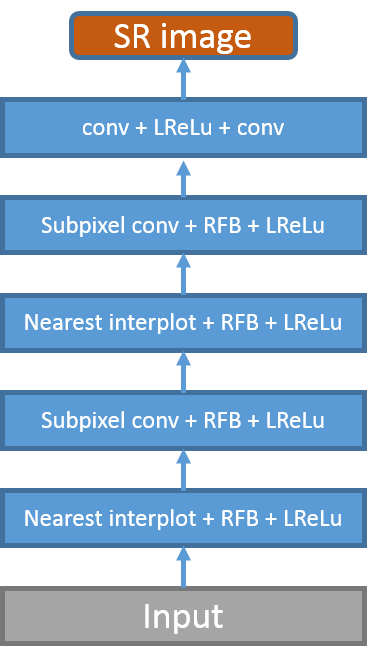}
\caption{Upsampling module.}
\label{fig:upsampling}
\end{figure}

The output $x_{RRFDB}$ of Trunk-RFB module is fed to a single RFB block and the upsampling module. In the upsampling phase, we alternately use Nearest Neighborhood Interpolation and  Sub-pixel Convolution\cite{shi2016real} shown in Fig. \ref{fig:upsampling}. The output of upsampling module can formulated as equation \eqref{eq:upsample}.where $f_{RFB}$ means the function of RFB, $f_{inter}$ means the function of Nearest Neighborhood Interpolation, $f_{sub}$ means the function of Sub-pixel Convolution.
\begin{equation}
    x=f_{sub(}f_{inter}(f_{sub}(f_{inter}(f_{RFB}(x_{RRFDB}))))) \label{eq:upsample}
\end{equation}
 Final convolution module consists of two layers of convolution with kernel size $3\times3$. Use $f_{c1}$ and $f_{c2}$ represent the functions of final two convolution layers, the final super resolution results can be given as equation \eqref{eq:result}.
\begin{equation}
    I_{SR}=f_{c2}(f_{c1}(x)) \label{eq:result}
\end{equation}

\subsection{Multi-scale Receptive Fields Block and Upsampling Module}
For perceptual extreme super resolution task, RFB-ESRGAN proposed to extract  multi-scale receptive fields feature for restoring details of the SR images. For this purpose, we need to assemble vary sizes of convolution filter into the generative network, such as $1 \times 1$, $3 \times 3$, $5 \times 5$. But large convolution kernel will greatly increase the time complexity of the model, it is needed to use small filters instead of large filters. In our work, we introduce the Receptive Fields Block (RFB) \cite{liu2018receptive} to assemble the RFB-ESRGAN. RFB has been proposed to strengthen the deep features learned from lightweight CNN models. Specifically, RFB makes use of multi-branch pooling with varying kernels corresponding to reception fields of different sizes, applies dilated convolution layers to control their eccentricities, and reshapes them to generate final representation. Here, the RFB is used in RRFDBs to remain the deep rich features for restoring the details of super resolution image.

\begin{figure}[htbp]
\centering 
\includegraphics[width = 0.45\textwidth]{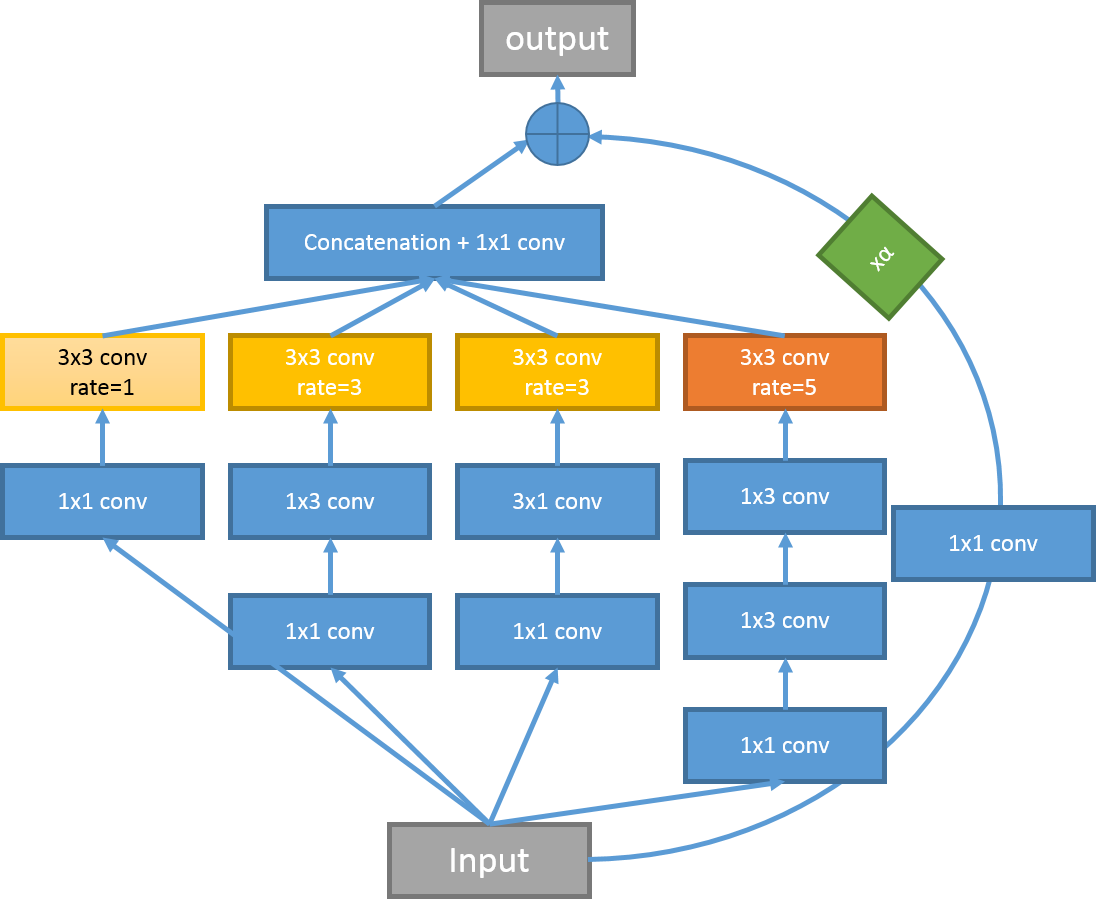}
\caption{Receptive Field Block (RFB).}
\label{fig:RFB}
\end{figure}

In RFB-ESRGAN, the trunk-RFB is stacked of 8 Residual of Receptive Field Dense Blocks (RRFDBs), and each RRFDB contains 5 RFBs (Fig. \ref{fig:RRFDB}). The composition structure of RFB is shown in Fig. \ref{fig:RFB}. RFB highlights the relationship between receptive filed size and eccentricity in a daisy-shape configuration, where bigger weights are assigned to the positions nearer to the center by smaller kernels, claiming that they are more important than the farther ones. This makes RFB more effect on simulating the human visual system than the other multi-scale receptive fields methods like the Inception family \cite{szegedy2015going}, ASPP\cite{chen2017rethinking}, and Deformable CNN \cite{dai2017deformable}. In the RFB, instead of large kernels such as $3 \times 3, 5 \times 5$, it uses the combination of small kernels ($1 \times 1, 1 \times 3, 3 \times 1$), which can effectively reduce the amount of parameters and time complexity. Besides, such substitutions enable RFB to extract very detailed features especially line features, such as hair, skin texture, edge, etc. This makes RFB exactly what we need for extracting multi-scale features and minimizing time complexity at the same time. The most important reason to use RFB is the ability of extracting the very detailed features, which is exactly what is needed in the field of image reconstruction.

To make RFB suitable for our RFB-ESRGAN, we drop all the batch normalization layers in RFB. In addition, we use Leaky Relu instead of Relu as the activation function of the whole RFB, while the activation functions in each branch are still Relu.

In the upsampling phase, instead of only use Nearest Neighborhood Interpolation (NNI) or Sub-pixel Convolution (SPC) \cite{shi2016real}, we alternately use NNI and SPC. NNI performs spatial transformation on input features, and the RFB after NNI makes the results of NNI's spatial transformation fully affect on depth. SPC makes depth to space transformation, and the RFB after SPC makes the results of SPC's depth to space transformation fully affect on space. Use them alternately will improve the information communication between space and depth. Also, the use of SPC will reduce the amount of parameters and time complexity.

\subsection{Loss Function} \label{sec:loss}
We apply GAN loss that used in ESRGAN \cite{wang2018esrgan} on RFB-ESRGAN, which results in the following loss for generative network and discriminator network. Generative loss function of RFB-ESRGAN contains three terms:  VGG loss which has been successfully applied on other tasks such as image synthesis and style transfer. The purpose of VGG loss here is encouraging our network to restore the high-frequency content for perceptually satisfaction. We use the pretrained VGG model to extract the feature representation of $I^{SR}$ and $I^{HR}$, $I^{SR}$ denotes the images generated by RFB-ESRGAN,  $I^{HR}$ denotes the ground truth high resolution images. Adversarial loss $L_{adv}$ for encouraging our network to favor solutions that reside on the manifold of natural images. Pixel loss $L_{pix}$ used to restrict the generation of too much high-frequency content. Use $\mathfrak D$ denotes the training dataset, $D(.)$ describes the discriminator network function, $G(.)$ describes the generative network function, and $\|.\|$ represents L1 loss. $I^{SR}$ can be formulated as equation \eqref{eq:sr}.
\begin{equation} \label{eq:sr}
\begin{split}
I^{SR}=G(I^{LR})\|
\end{split}
\end{equation}

Here $I^{LR}$ describes the input low resolution image. Pixel loss is the manhattan distance between reconstructed image $I^{SR}$ and the reference ground truth image $I^{HR}$, shown as equation \eqref{eq:pix_loss}.

\begin{equation} \label{eq:pix_loss}
\begin{split}
 L_{pix} = &\sum_{\mathfrak D}\|I^{SR}, I^{HR}\|
\end{split}
\end{equation}

VGG loss is the manhattan distance between the VGG feature representations of a reconstructed image $I^{SR}$ and the reference ground truth image $I^{HR}$, shown as equation \eqref{eq:vgg_loss}.

\begin{equation} \label{eq:vgg_loss}
 L_{VGG} = \sum_{\mathfrak D}\|VGG_{conv34}(I^{SR}), VGG_{conv34}(I^{HR})\|
\end{equation}

Where $VGG_{conv34}$ represents the feature map of $34$th layer in pretrained VGG model. Use $\Delta(.)$ represents the difference between the realistic degree of reconstructed image $I^{SR}$ and reference ground truth image $I^{HR}$, the difference between $I^{SR}$ and $I^{HR}$ shown as \eqref{eq:delta}. The adversarial loss can be formulated as equation \eqref{eq:adv_loss}.
\begin{equation} \label{eq:delta}
\begin{split}
\Delta_{Real} = \sigma(D(I^{HR}) - E(D(I^{SR})))\\ 
\Delta_{Fake} = \sigma(D(I^{SR}) - E(D(I^{HR})))\\ 
\end{split}
\end{equation}
Where $\sigma$ is the sigmoid function and $E[.]$ represents the average operation of all data in a mini-batch.

\begin{equation} \label{eq:adv_loss}
\begin{split}
L_{adv} = -E[log(1-\Delta_{Real})] -E[ log(\Delta_{Fake})]\\ 
\end{split}
\end{equation}

With pixel loss, VGG loss,  and adversarial loss, we can formulate the generative loss of RFB-ESRGAN shown as equation \eqref{eq:generative_loss}.
\begin{equation} \label{eq:generative_loss}
\begin{split}
L_{G} = \lambda L_{pix} +  L_{VGG} + \eta L_{adv}\\ 
\end{split}
\end{equation}

Discriminator loss function of RFB-ESRGAN contains two terms: Real Loss $L_{Real}$ for encouraging the real image is more realistic than fake image, shown as \eqref{eq:real_loss}. Fake loss $L_{Fake}$ for encouraging the fake image is less realistic than real image, shown as equation \eqref{eq:fake_loss}. 

\begin{equation} \label{eq:real_loss}
\begin{split}
L_{Real} = -E[log(\Delta_{Real})]\\ 
\end{split}
\end{equation}

\begin{equation} \label{eq:fake_loss}
\begin{split}
L_{Fake} = -E[1-log(\Delta_{Fake})]\\ 
\end{split}
\end{equation}

With the real loss $L_{Real}$ and fake loss $L_{Fake}$, the loss function of discriminator can be formulated as equation \eqref{eq:discriminator_loss}.
\begin{equation} \label{eq:discriminator_loss}
\begin{split}
L_{D} = L_{Real} + L_{Fake}\\ 
\end{split}
\end{equation}

\subsection{Model Ensemble}
Different from ESRGAN \cite{wang2018esrgan}, which fuses the parameters of PSNR-oriented model $G_{PSNR}$ and GAN-based model $G_{GAN}$. In order to extremely improve the perceptual performance of the reconstructed image, we fuse the model without any PSNR-oriented model. The final model is ensemble of 10 GAN-based models with the best perceptual performance among all recorded models in GAN training stage. We fuse all the corresponding parameters of top 10 models to derive an ensemble model $G_{Ensemble}$, whose parameters are:
\begin{equation} \label{eq:ensemble}
\begin{split}
\theta_{G}^{Ensemble}=\frac{1}{N} \sum_i^N(\theta_G^{GAN})
\end{split}
\end{equation}

where $\theta_{G}^{Ensemble}$ represents the parameters of $G_{Ensemble}$, $\theta_G^{GAN}$ represents the parameters of $G_{GAN}$, and $N$ is set as $10$ for NTIRE 2020 Perceptual Extreme Super-Resolution Challenge. The final ensemble model $G^{Ensemble}$ can effectively reduce the noise of reconstructed images and be more robust for different test images. We also attempt to fuse the models with more GAN-based models. For instance, use 20 or 40 best GAN-based models for ensemble. We find that, the ensemble model with more GAN-based models can reduce the noise of reconstructed images a little more. However, it doesn't further improve the model's perceptual performance of ensemble model. Instead, with more models for ensemble has a negative impact on perceptual performance. We balanced the performance of different numbers of fusion models, and finally chose to use 10 models for ensemble.

\begin{figure*}
    \centering
    \subfloat[1608 from DIV8K]{\includegraphics[width=0.19\textwidth]{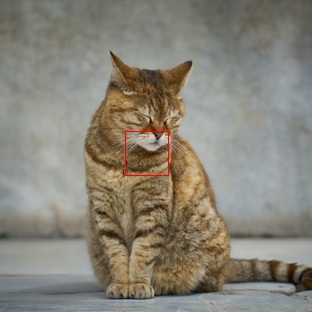}}\hspace{0.005in}
    \subfloat{\includegraphics[width=0.19\textwidth]{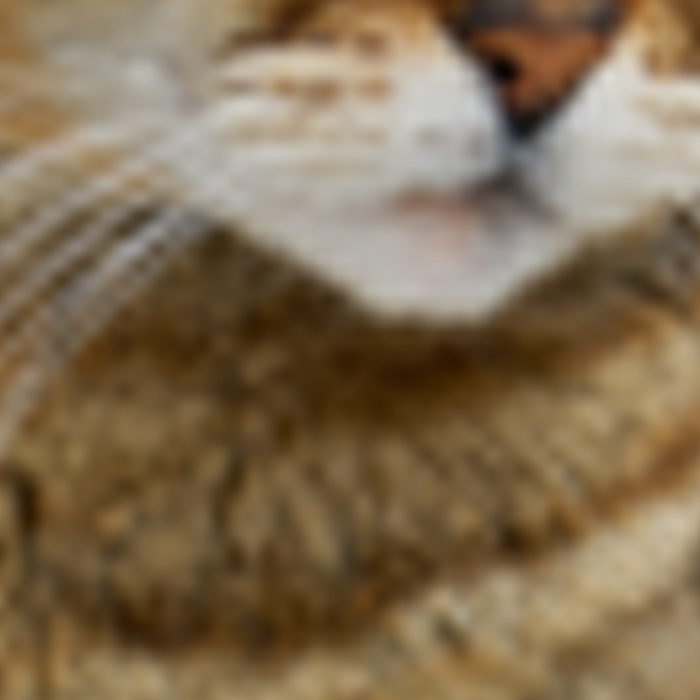}}\hspace{0.005in}
    \subfloat{\includegraphics[width=0.19\textwidth]{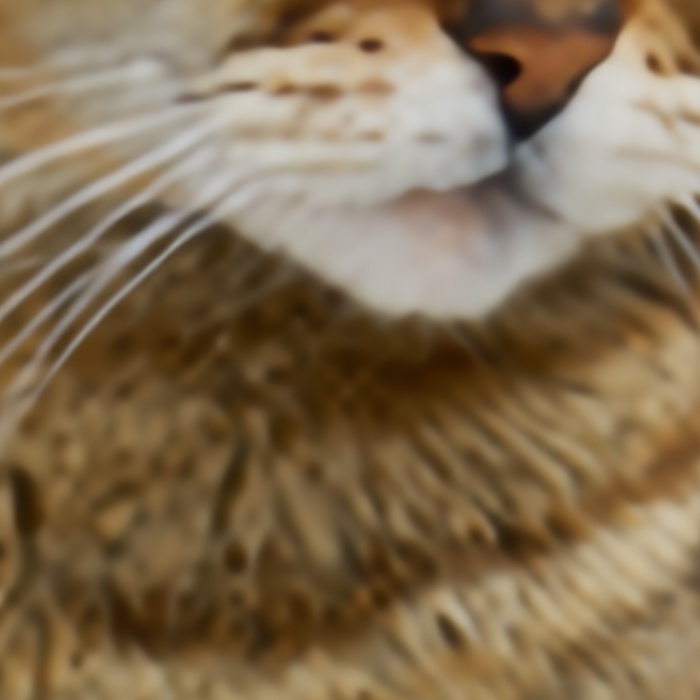}}\hspace{0.005in}
    \subfloat{\includegraphics[width=0.19\textwidth]{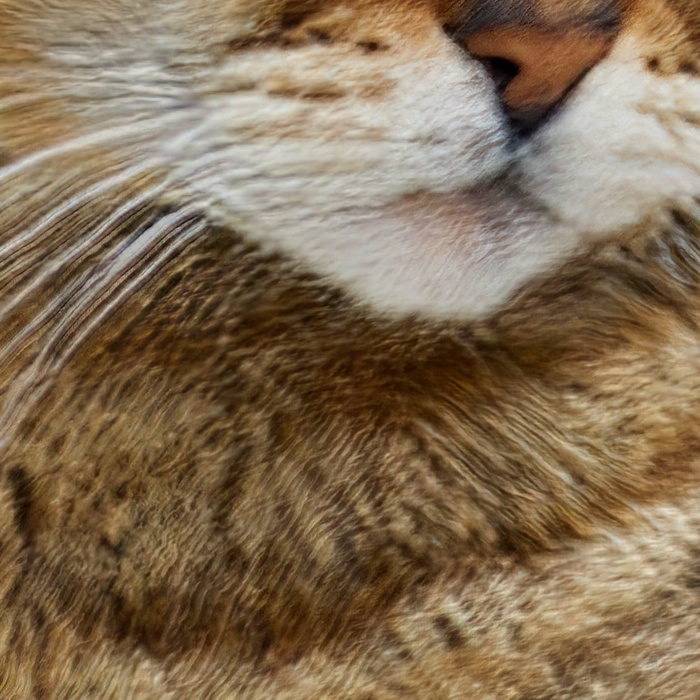}}\hspace{0.005in}
    \subfloat{\includegraphics[width=0.19\textwidth]{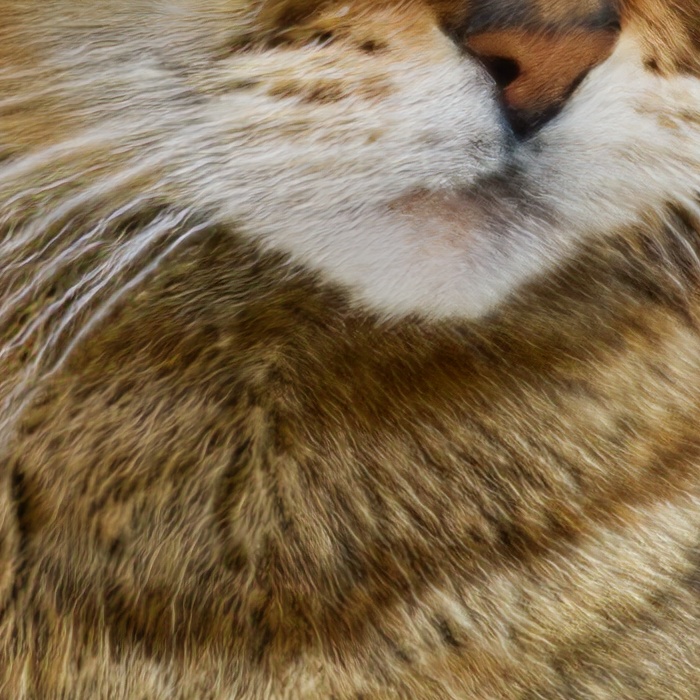}} \vspace{-0.15in}
    \subfloat[1619 from DIV8K\protect\\LR]{\includegraphics[width=0.19\textwidth]{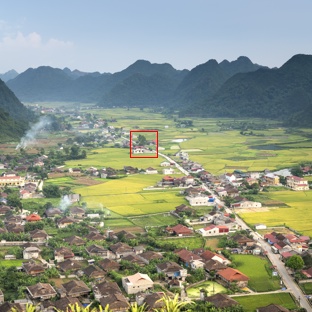}}\hspace{0.005in}
    \subfloat[ \protect\\bicubic]{\includegraphics[width=0.19\textwidth]{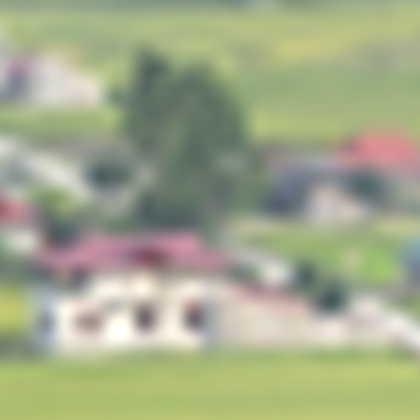}}\hspace{0.005in}
    \subfloat[ \protect\\RCAN]{\includegraphics[width=0.19\textwidth]{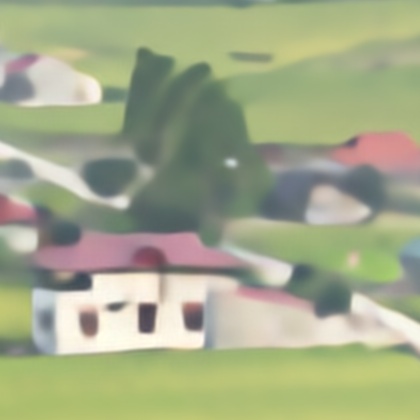}}\hspace{0.005in}
    \subfloat[ \protect\\ESRGAN]{\includegraphics[width=0.19\textwidth]{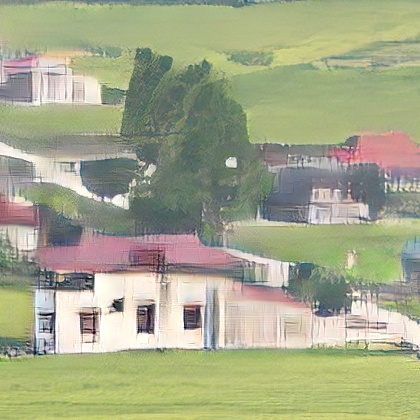}}\hspace{0.005in}
    \subfloat[ \protect\\RFB-ESRGAN]{\includegraphics[width=0.19\textwidth]{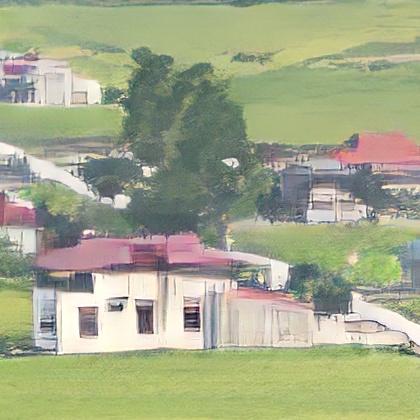}}\vspace{-0.15in}

    \caption{Qualitative results of RFB-ESRGAN. RFB-ESRGAN produces more natural textures, e.g., animal fur, building structure and plant texture.}
\label{fig:result1}
\end{figure*}

\section{Experiments}
\subsection{Training Details}
For NTIRE 2020 Perceptual Extreme Super-Resolution Challenge, all experiments are performed with a scaling factor of $\times$16 between LR and HR images. We obtained the corresponding LR images via default setting (bicubic interpolation) of Matlab function "imresize" with scale factor 16. The mini-batch size is set as 16. The spatial size of cropped HR patch is $512 \times 512$, and spatial size of corresponding input LR image is $32 \times 32$. 

The training process can be divided into two stages. First stage, a PSNR-oriented model with $L1$ loss is trained. The learning rate is initialized as $2 \times 10^{-4}$ and decayed by a factor of $2$ every $2.5 \times 10^5$ of mini-batch steps. Second stage (GAN-based training stage), after fully trained of PSNR-oriented model, generative network is initialized with the parameters of pre-trained PSNR-oriented model and trained using the generative loss function \eqref{eq:generative_loss} and adversarial loss function \eqref{eq:adv_loss}. In the generative loss function, $\lambda$ is set as 10 and $\eta$ is set as $5e^{-3}$. The learning rate is set as $1e^{-4}$ and halved at $[50k,100k,200k,300k]$ iterations. During the GAN-based training stage, parameters of generative network is recorded every 5000 iteration. 

For optimization, we use Adam \cite{kingma2014adam} with $\beta_1=0.9$ and $\beta_2=0.99$. The generative network and discriminator network are alternately updated. we implement our models with Pytorch framework and train them using Tesla V100 GPUs. There are 20.5M parameters in RFB-ESRGAN model, and it costs 0.82s using one Tesla V100 GPU for processing per image with 128x128 pixels.

\subsection{Data}
NTIRE 2020 Perceptual Extreme Super-Resolution Challenge has provided DIV8K dataset \cite{gu2019div8k} for training, which includes 1,500 HR images with high resolution vary from 2K to 8K, we use 1,400 images for training and the rest 100 images for validation. In order to enrich our training dataset, we added other datasets, including 800 images from DIV2k dataset \cite{agustsson2017ntire}, 2,650 images from Flickr2K \cite{timofte2017ntire} dataset and 785 images from OST dataset \cite{wang2018recovering}.

Our models are trained with RGB channels. For data augmenting, the training images are random flipped with horizontal and random rotated with 90 degree. The result models are evaluated on DIV8K dataset provided by NTIRE 2020 Perceptual Extreme Super-Resolution Challenge.

\subsection{Qualitative Results}
We have compared our final models on DIV8K with PSNR-oriented method RCNN, and also with perceptual-driven approach ESRGAN. Because the original RCNN and ESRGAN didn't adjust to $\times$16 scale super resolution task, we finetuned them on datasets same as the proposed RFB-ESRGAN. We present some representative qualitative results in Fig. \ref{fig:result1}. 

From Fig. \ref{fig:result1}, we can observe that, our proposed RFB-ESRGAN outperforms previous approaches in both similar to ground truth and details. For instance, RFB-ESRGAN can produce sharper and clearer textures than PSNR-oriented method RCAN. The PSNR-oriented methods always tend to produce smooth and blurry SR images, which is not friendly to human visual perception. RFB-ESRGAN is capable of generating sharper and more natural details than ESRGAN. The fur textures of cat (see image 1608) are more real, the textures of plants and buildings (see image 1643) are more natural.

\begin{table}[htbp]
\centering
\begin{tabular}{llccc}
\hline \hline
\multicolumn{1}{l|}{Method}          & PSNR $\uparrow$   & SSIM $\uparrow$    & LPIPS $\downarrow$  & PI $\downarrow$  \\ \hline
\multicolumn{1}{l|}{BICUBIC}      & 24.67  & 0.59  & 0.656  & 11.29 \\
\multicolumn{1}{l|}{RCAN}        & \textbf{25.90}  & \textbf{0.62}  & 0.548  & 9.16 \\
\multicolumn{1}{l|}{ESRGAN}   & 23.98  & 0.53  & 0.351  & \textbf{3.89} \\
\multicolumn{1}{l|}{RFB-ESRGAN}  & 24.03  & 0.54  & \textbf{0.345}  & 4.27  \\
\hline\hline
\end{tabular}
\caption{The PSNR, SSIM, LPIPS and PI are calculated on the center 1,000x1,000 subimages of 1,401-1,500 images from the DIV8K.}
\label{tab:validation1}
\end{table}

We also compare the results on 1,401-1,500 images from DIV8K, which haven't been used for training. PSNR, SSIM, LPIPS and PI were calculated to evaluate the sharpness and fidelity of results. The results are shown in Tab. \ref{tab:validation1}, in which the arrows indicate if high $\uparrow$ or low $\downarrow$ values are desired. Besides, our solution RFB-ESRGAN won the NTIRE 2020 Perceptual Extreme Super-Resolution Challenge according to preliminary results. We present the top 6 results from the Challenge in Tab. \ref{tab:challenge}, more information on the evaluation and competing methods can be found in the challenge report \cite{zhang2020ntire}.

\begin{table}[htbp]
\centering
\resizebox{0.5\textwidth}{!}{
\begin{tabular}{llccc}
\hline \hline
\multicolumn{1}{l|}{Team}          & PSNR $\uparrow$   & SSIM $\uparrow$    & LPIPS $\downarrow$  & PI $\downarrow$   \\ \hline
\multicolumn{1}{l|}{Our Team}      & 23.38  & 0.5504  & \textbf{0.348}  & 3.97 \\
\multicolumn{1}{l|}{CIPLAB}        & 22.77  & 0.5251  & 0.352  & \textbf{3.76} \\
\multicolumn{1}{l|}{HiImageTeam}   & 23.53  & 0.5624  & 0.368  & 4.38 \\
\multicolumn{1}{l|}{Winner AIM19}  & 24.52  & 0.5800  & 0.418  & 6.28      \\
\multicolumn{1}{l|}{ECNU}          & \textbf{25.56}  & \textbf{0.6336}  & 0.497  & 8.10 \\
\multicolumn{1}{l|}{SIA}           & 22.86  & 0.5896  & 0.434  & 5.81 \\ 
\hline\hline
\end{tabular}}
\caption{Results of NTIRE 2020 perceptual extreme SR challenge. The PSNR, SSIM, LPIPS and PI are calculated on the center 1,000x1,000 subimages of the DIV8K test images \cite{zhang2020ntire}.}
\label{tab:challenge}
\end{table}

\subsection{Ablation Study}
\begin{figure*}[htbp]
\centering
    \begin{minipage}[b]{0.846\textwidth}
    \centering
        \begin{tabular}{>{\centering\arraybackslash}p{0.14\textwidth}>{\centering\arraybackslash}p{0.14\textwidth}>{\centering\arraybackslash}p{0.14\textwidth}>{\centering\arraybackslash}p{0.14\textwidth}>{\centering\arraybackslash}p{0.14\textwidth}>{\centering\arraybackslash}p{0.14\textwidth}}
        1      & 2 & 3 & 4 & 5 & 6 \\
        Ensemble & \xmark  & \xmark  & \xmark  & \xmark  &  \checkmark \\
        RFB    & \checkmark    & \checkmark  & \xmark  & \checkmark  & \checkmark  \\
        SPC    & \checkmark      & \xmark & \checkmark  & \checkmark  & \checkmark \\
        NNI    & \xmark  & \checkmark  & \checkmark  & \checkmark  & \checkmark \\
        \end{tabular}
    \end{minipage}
    \begin{minipage}[b]{\textwidth}
    \centering
    \subfloat[1601 from DIV8K]{\includegraphics[width=0.14\textwidth]{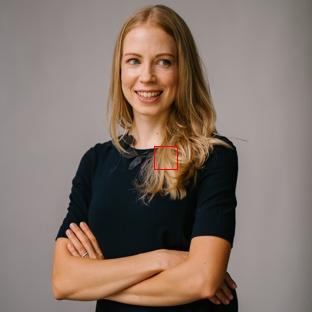}}\hspace{0.001\textwidth}
    \subfloat{\includegraphics[width=0.14\textwidth]{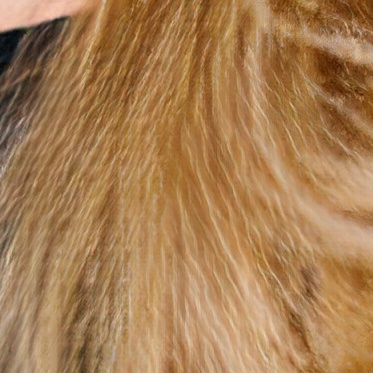}}\hspace{0.001\textwidth}
    \subfloat{\includegraphics[width=0.14\textwidth]{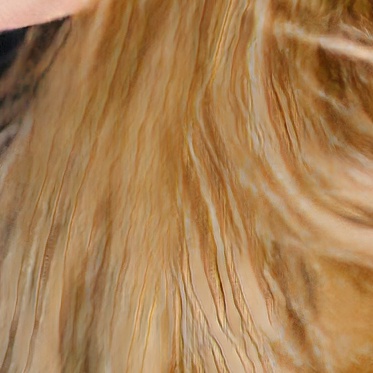}}\hspace{0.001\textwidth}
    \subfloat{\includegraphics[width=0.14\textwidth]{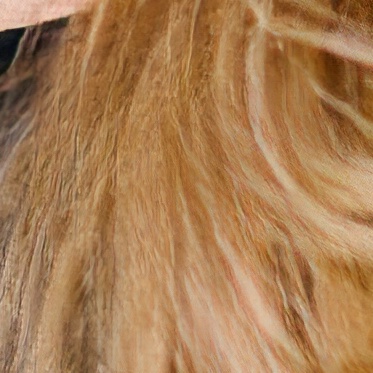}}\hspace{0.001\textwidth}
    \subfloat{\includegraphics[width=0.14\textwidth]{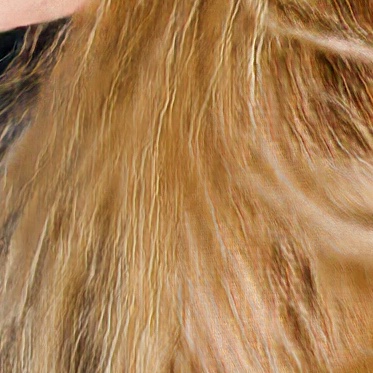}}\hspace{0.001\textwidth}
    \subfloat{\includegraphics[width=0.14\textwidth]{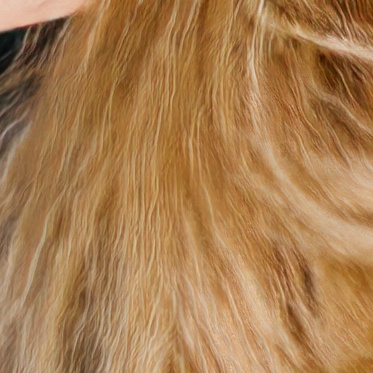}}
    \vspace{-0.15in}

    \subfloat[1608 from DIV8K]{\includegraphics[width=0.14\textwidth]{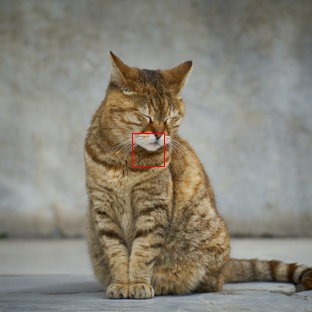}}\hspace{0.001\textwidth}
    \subfloat{\includegraphics[width=0.14\textwidth]{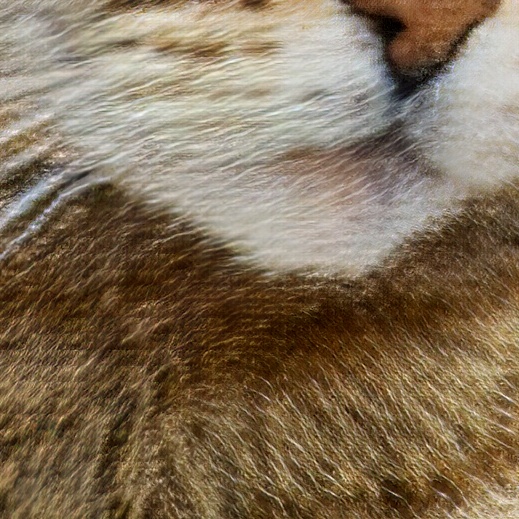}}\hspace{0.001\textwidth}
    \subfloat{\includegraphics[width=0.14\textwidth]{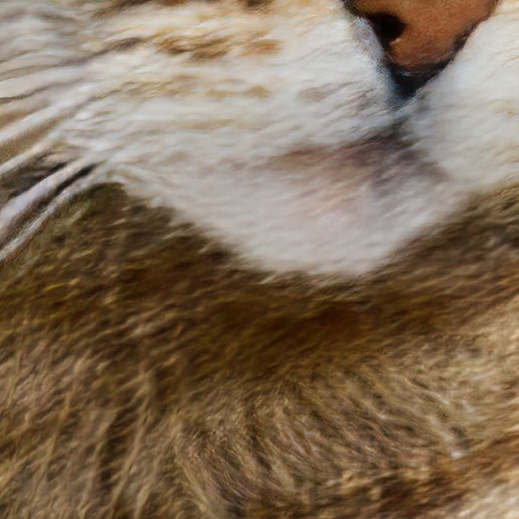}}\hspace{0.001\textwidth}
    \subfloat{\includegraphics[width=0.14\textwidth]{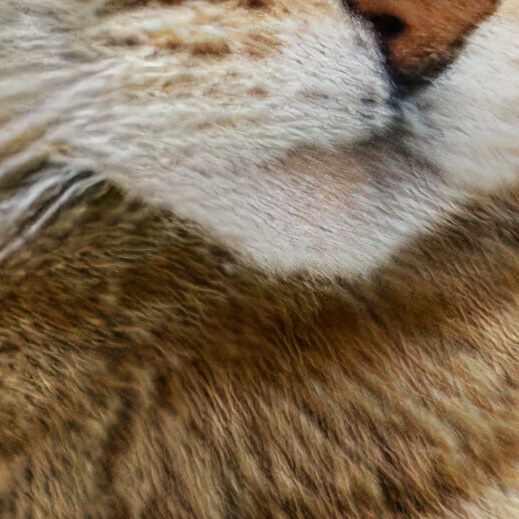}}\hspace{0.001\textwidth}
    \subfloat{\includegraphics[width=0.14\textwidth]{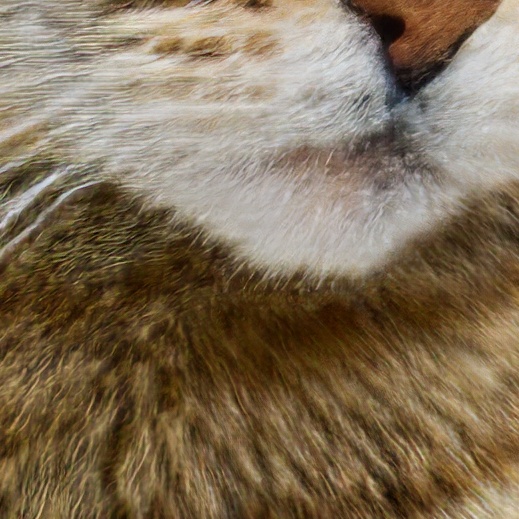}}\hspace{0.001\textwidth}
    \subfloat{\includegraphics[width=0.14\textwidth]{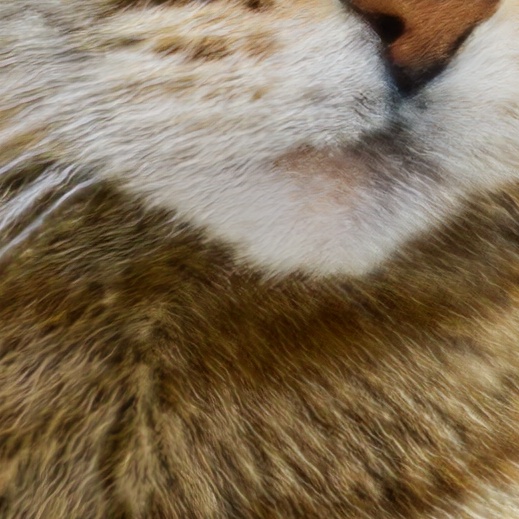}}
    \vspace{-0.15in}
    
    \subfloat[1617 from DIV8K]{\includegraphics[width=0.14\textwidth]{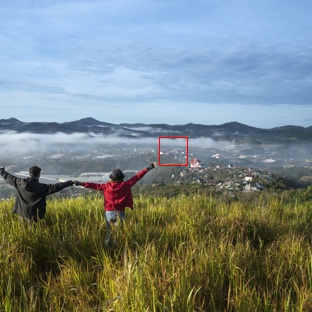}}\hspace{0.001\textwidth}
    \subfloat{\includegraphics[width=0.14\textwidth]{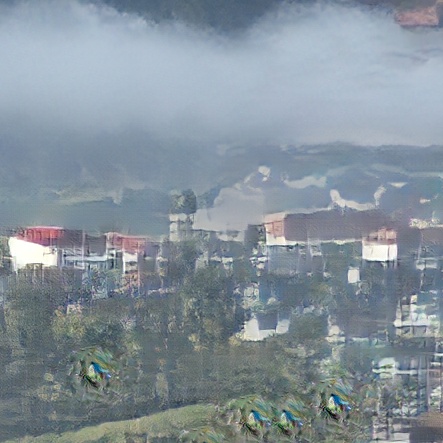}}\hspace{0.001\textwidth}
    \subfloat{\includegraphics[width=0.14\textwidth]{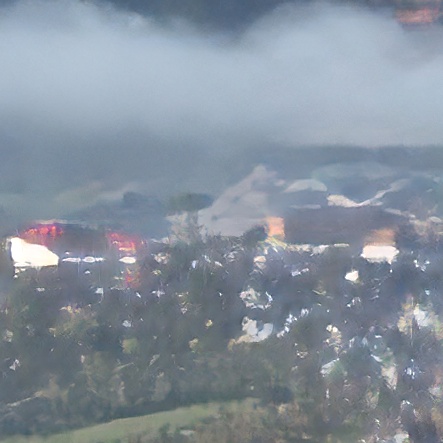}}\hspace{0.001\textwidth}
    \subfloat{\includegraphics[width=0.14\textwidth]{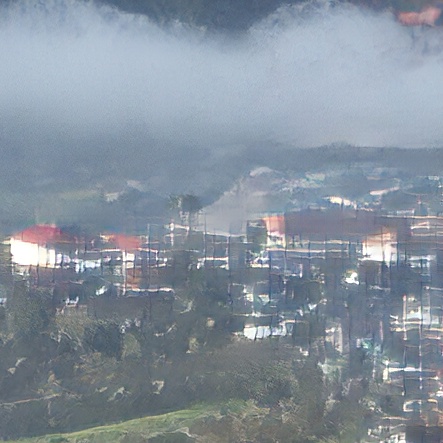}}\hspace{0.001\textwidth}
    \subfloat{\includegraphics[width=0.14\textwidth]{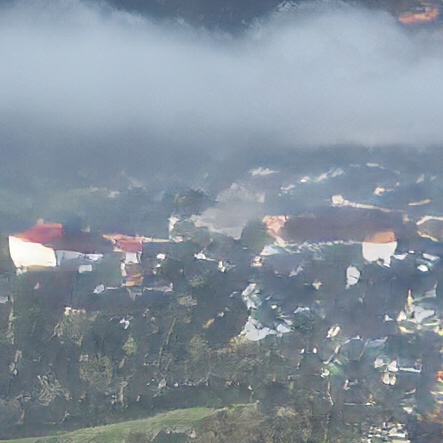}}\hspace{0.001\textwidth}
    \subfloat{\includegraphics[width=0.14\textwidth]{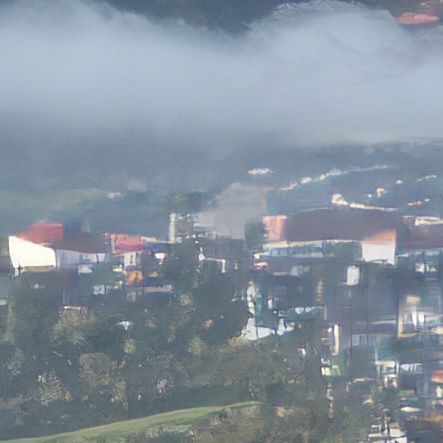}}
    \vspace{-0.15in}

    \subfloat[1643 from DIV8K]{\includegraphics[width=0.14\textwidth]{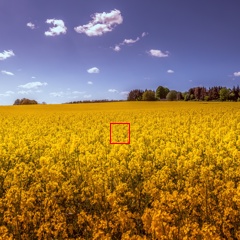}}\hspace{0.001\textwidth}
    \subfloat{\includegraphics[width=0.14\textwidth]{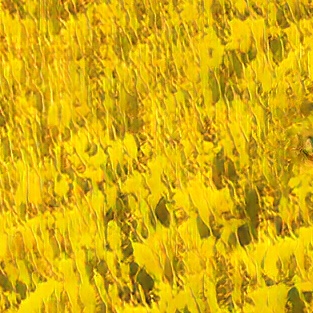}}\hspace{0.001\textwidth}
    \subfloat{\includegraphics[width=0.14\textwidth]{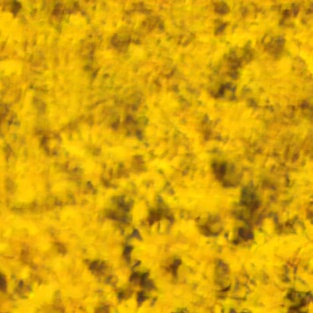}}\hspace{0.001\textwidth}
    \subfloat{\includegraphics[width=0.14\textwidth]{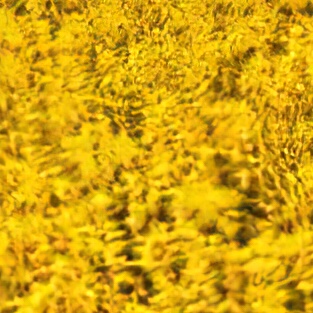}}\hspace{0.001\textwidth}
    \subfloat{\includegraphics[width=0.14\textwidth]{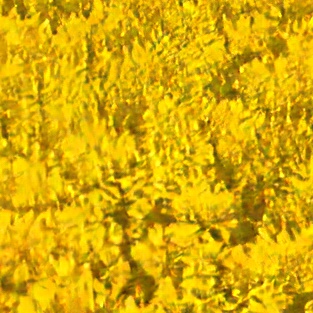}}\hspace{0.001\textwidth}
    \subfloat{\includegraphics[width=0.14\textwidth]{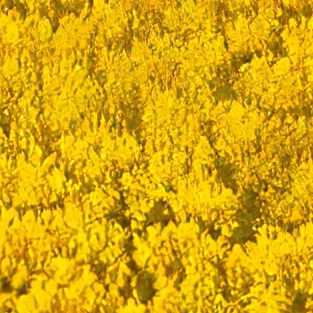}}
    \vspace{-0.15in}
    
    \subfloat[1645 from DIV8K]{\includegraphics[width=0.14\textwidth]{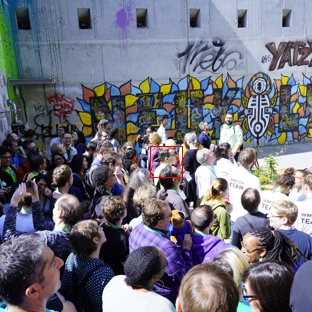}}\hspace{0.001\textwidth}
    \subfloat{\includegraphics[width=0.14\textwidth]{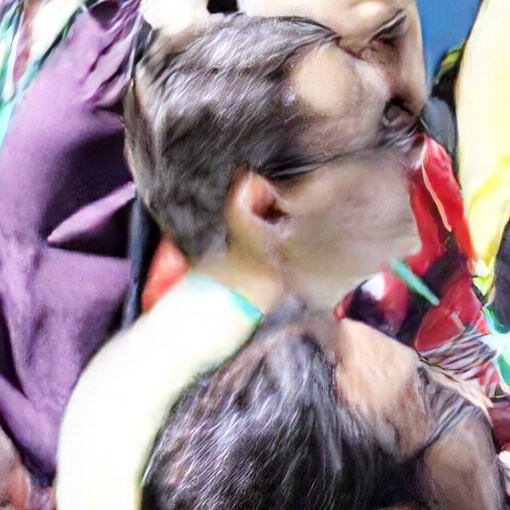}}\hspace{0.001\textwidth}
    \subfloat{\includegraphics[width=0.14\textwidth]{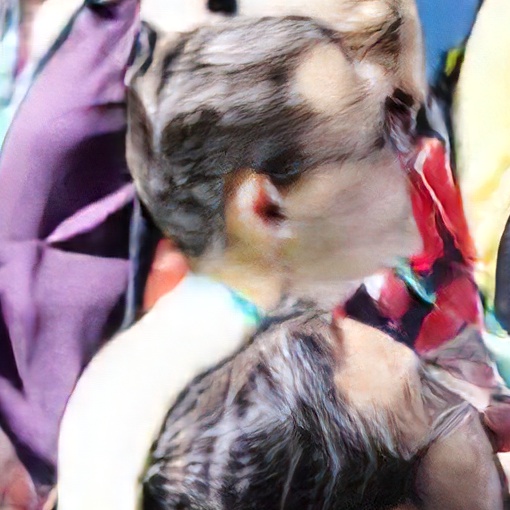}}\hspace{0.001\textwidth}
    \subfloat{\includegraphics[width=0.14\textwidth]{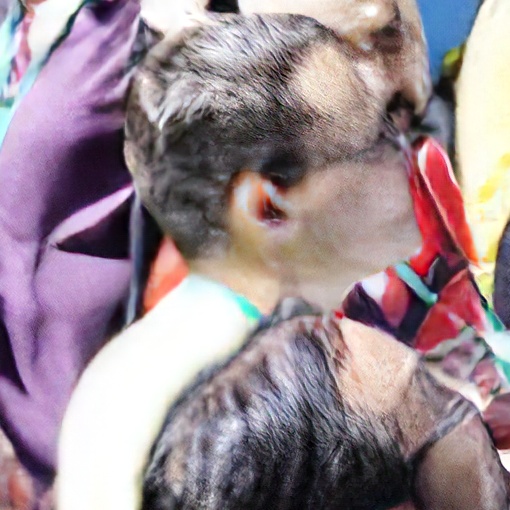}}\hspace{0.001\textwidth}
    \subfloat{\includegraphics[width=0.14\textwidth]{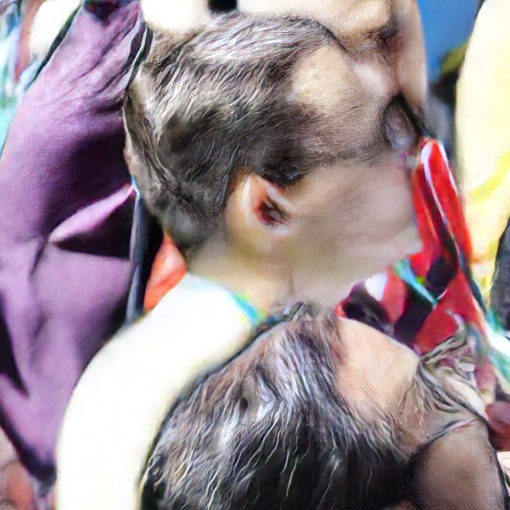}}\hspace{0.001\textwidth}
    \subfloat{\includegraphics[width=0.14\textwidth]{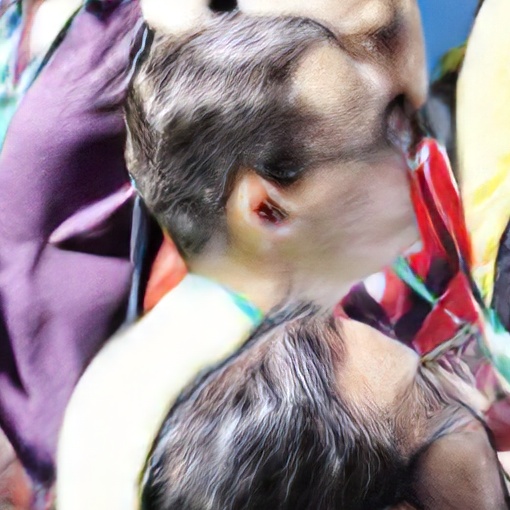}}
    \vspace{-0.15in}
    
    \end{minipage}
\caption{Overall visual comparisons for showing the effects of each component in RFB-ESRGAN. SPC means Sub-pixel convolution and NNI means Nearest Neighbor Interpolation, which are used in upsampling stage.}
\label{fig:ablation}
\end{figure*}

In order to study the effects of each component of the proposed RFB-ESRGAN, we remove the different components of RFB-ESRGAN to measure the influence of it. The overall visual comparison is shown in Fig. \ref{fig:ablation}. Each column images represents the super resolution results of the model with configurations shown in the top. Among them, Configurations of $2^{nd}$ column represents the model use only Subpixel Convolution for upsampling, $3^{nd}$ column represents the model use only Nearest Neighbor Interpolation for upsampling, $4^{nd}$ colunn represents the model Alternately use Subpixel Convolution and Nearest Neighbor Interpolation for upsampling. Detailed of ablation study is provided as follows.

\textbf{RFB}. In order to prove the effect of RFB, we remove all RFBs in the model while keep the entire structure of the model unchanged. From some cases of $4^{nd}$ column, we can observe that the textures of hair from people and fur from cat are too rough, and some with wrong direction. While the results of RFB-ESRGAN in $5^{nd}$ column achieve fine and smooth hair and fur. 

\textbf{Methods for Upsampling}. We have Alternately used Nearest Neighbor Interpolation (NNI) and Subpixel Convolution (SPC) in upsampling stage, shown in Fig. \ref{fig:upsampling}. In order to demonstrate the effect of this upsampling methods, we test the upsampling methods of using only NNI in $3^{nd}$ column and using only SPC in $2^{nd}$ column. As shown in the $3^{nd}$ column, results of the method with only NNI are more blurry than the other upsampling methods. While using only SPC, the textures of some cases are too sharp and not natural (see image 1608 and 1643 in $2^{nd}$ column), and also some unreal artifacts have been generated (see image 1617 in $2^{nd}$ column). It can be observed our upsampling method yields the most clear and realistic results. 

\textbf{Ensemble}. To evaluate the effect of model ensemble, we compare the SR results with model ensemble and without model ensemble. From $5^{nd}$ column, we can observe that the results without ensemble have obvious noise though the textures are sharper and clear. While  most noises can be eliminated by model ensemble as shown in $6^{nd}$. The hair textures become more natural (see image 1601 and image 1645), and the noise is suppressed to some extent (see image 1617 and 1643).

Besides, we have calculated PSNR, SSIM, LPIPS and PI on the results of 1,401-1,500 images form DIV8K, which haven't been used for training. The results are shown in Tab. \ref{tab:validation2}. The configuration of each $n^{nd}$ column is as shown as Fig. \ref{fig:ablation}.

\begin{table}[htbp]
\centering
\begin{tabular}{llccc}
\hline \hline
\multicolumn{1}{l|}{Method}          & PSNR $\uparrow$   & SSIM $\uparrow$    & LPIPS $\downarrow$  & PI $\downarrow$   \\ \hline

\multicolumn{1}{l|}{$2^{nd}$ column} & 23.40  & 0.50  & 0.370  & \textbf{3.73} \\
\multicolumn{1}{l|}{$3^{nd}$ column} & \textbf{24.09}  & \textbf{0.54}  & 0.363  & 4.18 \\
\multicolumn{1}{l|}{$4^{nd}$ column} & 23.60  & 0.52  & 0.365  & 3.93 \\
\multicolumn{1}{l|}{$5^{nd}$ column} & 23.60  & 0.52  & 0.357  & 3.92      \\
\multicolumn{1}{l|}{$6^{nd}$ column} & 24.03  & \textbf{0.54}  & \textbf{0.345}  & 4.27      \\
\hline\hline
\end{tabular}
\caption{The PSNR, SSIM, LPIPS and PI are calculated on the center 1,000x1,000 subimages of 1,401-1,500 images from the DIV8K.}
\label{tab:validation2}
\end{table}

\section{Conclusion}
We proposed RFB-ESRGAN for single image extreme perceptual super resolution problem. For $\times$16 scale super resolution, we proposed using multi-scale receptive fields for extracting multi-scale features of LR image. In addition, we proposed using small convolution kernels to extract detailed features of input image for reconstructing the detailed features of SR image. We have also proposed using nearest interpolation and sub-pixel convolution alternately for improving the information exchange between spacial and depth, and reducing the amount of parameters in upsampling stage. Our experiments and the results of NTIRE 2020 Perceptual Extreme Super-Resolution Challenge have demonstrate the effectiveness of our solution for perceptual extreme super-resolution.

{\small
\bibliographystyle{ieee_fullname}
\bibliography{egbib}

\begin{thebibliography}{10}\itemsep=-1pt

\bibitem{agustsson2017ntire}
Eirikur Agustsson and Radu Timofte.
\newblock Ntire 2017 challenge on single image super-resolution: Dataset and
  study.
\newblock In {\em Proceedings of the IEEE Conference on Computer Vision and
  Pattern Recognition Workshops}, pages 126--135, 2017.

\bibitem{chang2004super}
Hong Chang, Dit-Yan Yeung, and Yimin Xiong.
\newblock Super-resolution through neighbor embedding.
\newblock In {\em Proceedings of the 2004 IEEE Computer Society Conference on
  Computer Vision and Pattern Recognition, 2004. CVPR 2004.}, volume~1, pages
  I--I. IEEE, 2004.

\bibitem{chen2017rethinking}
Liang-Chieh Chen, George Papandreou, Florian Schroff, and Hartwig Adam.
\newblock Rethinking atrous convolution for semantic image segmentation.
\newblock {\em arXiv preprint arXiv:1706.05587}, 2017.

\bibitem{dai2017deformable}
Jifeng Dai, Haozhi Qi, Yuwen Xiong, Yi Li, Guodong Zhang, Han Hu, and Yichen
  Wei.
\newblock Deformable convolutional networks.
\newblock In {\em Proceedings of the IEEE international conference on computer
  vision}, pages 764--773, 2017.

\bibitem{dai2009softcuts}
Shengyang Dai, Mei Han, Wei Xu, Ying Wu, Yihong Gong, and Aggelos~K
  Katsaggelos.
\newblock Softcuts: a soft edge smoothness prior for color image
  super-resolution.
\newblock {\em IEEE Transactions on Image Processing}, 18(5):969--981, 2009.

\bibitem{dong2014learning}
Chao Dong, Chen~Change Loy, Kaiming He, and Xiaoou Tang.
\newblock Learning a deep convolutional network for image super-resolution.
\newblock In {\em European conference on computer vision}, pages 184--199.
  Springer, 2014.

\bibitem{dong2015image}
Chao Dong, Chen~Change Loy, Kaiming He, and Xiaoou Tang.
\newblock Image super-resolution using deep convolutional networks.
\newblock {\em IEEE transactions on pattern analysis and machine intelligence},
  38(2):295--307, 2015.

\bibitem{dong2016accelerating}
Chao Dong, Chen~Change Loy, and Xiaoou Tang.
\newblock Accelerating the super-resolution convolutional neural network.
\newblock In {\em European conference on computer vision}, pages 391--407.
  Springer, 2016.

\bibitem{duchon1979lanczos}
Claude~E Duchon.
\newblock Lanczos filtering in one and two dimensions.
\newblock {\em Journal of applied meteorology}, 18(8):1016--1022, 1979.

\bibitem{freeman2002example}
William~T Freeman, Thouis~R Jones, and Egon~C Pasztor.
\newblock Example-based super-resolution.
\newblock {\em IEEE Computer graphics and Applications}, 22(2):56--65, 2002.

\bibitem{gatys2015neural}
Leon~A Gatys, Alexander~S Ecker, and Matthias Bethge.
\newblock A neural algorithm of artistic style.
\newblock {\em arXiv preprint arXiv:1508.06576}, 2015.

\bibitem{goodfellow2014generative}
Ian Goodfellow, Jean Pouget-Abadie, Mehdi Mirza, Bing Xu, David Warde-Farley,
  Sherjil Ozair, Aaron Courville, and Yoshua Bengio.
\newblock Generative adversarial nets.
\newblock In {\em Advances in neural information processing systems}, pages
  2672--2680, 2014.

\bibitem{gu2019aim}
Shuhang Gu, Martin Danelljan, Radu Timofte, Muhammad Haris, Kazutoshi Akita,
  Greg Shakhnarovic, Norimichi Ukita, Pablo~Navarrete Michelini, Wenbin Chen,
  Hanwen Liu, et~al.
\newblock Aim 2019 challenge on image extreme super-resolution: Methods and
  results.
\newblock In {\em 2019 IEEE/CVF International Conference on Computer Vision
  Workshop (ICCVW)}, pages 3556--3564. IEEE, 2019.

\bibitem{gu2019div8k}
Shuhang Gu, Andreas Lugmayr, Martin Danelljan, Manuel Fritsche, Julien Lamour,
  and Radu Timofte.
\newblock Div8k: Diverse 8k resolution image dataset.
\newblock In {\em 2019 IEEE/CVF International Conference on Computer Vision
  Workshop (ICCVW)}, pages 3512--3516. IEEE, 2019.

\bibitem{howard2017mobilenets}
Andrew~G Howard, Menglong Zhu, Bo Chen, Dmitry Kalenichenko, Weijun Wang,
  Tobias Weyand, Marco Andreetto, and Hartwig Adam.
\newblock Mobilenets: Efficient convolutional neural networks for mobile vision
  applications.
\newblock {\em arXiv preprint arXiv:1704.04861}, 2017.

\bibitem{keys1981cubic}
Robert Keys.
\newblock Cubic convolution interpolation for digital image processing.
\newblock {\em IEEE transactions on acoustics, speech, and signal processing},
  29(6):1153--1160, 1981.

\bibitem{kim2016accurate}
Jiwon Kim, Jung Kwon~Lee, and Kyoung Mu~Lee.
\newblock Accurate image super-resolution using very deep convolutional
  networks.
\newblock In {\em Proceedings of the IEEE conference on computer vision and
  pattern recognition}, pages 1646--1654, 2016.

\bibitem{kingma2014adam}
Diederik~P Kingma and Jimmy Ba.
\newblock Adam: A method for stochastic optimization.
\newblock {\em arXiv preprint arXiv:1412.6980}, 2014.

\bibitem{ledig2017photo}
Christian Ledig, Lucas Theis, Ferenc Husz{\'a}r, Jose Caballero, Andrew
  Cunningham, Alejandro Acosta, Andrew Aitken, Alykhan Tejani, Johannes Totz,
  Zehan Wang, et~al.
\newblock Photo-realistic single image super-resolution using a generative
  adversarial network.
\newblock In {\em Proceedings of the IEEE conference on computer vision and
  pattern recognition}, pages 4681--4690, 2017.

\bibitem{lim2017enhanced}
Bee Lim, Sanghyun Son, Heewon Kim, Seungjun Nah, and Kyoung Mu~Lee.
\newblock Enhanced deep residual networks for single image super-resolution.
\newblock In {\em Proceedings of the IEEE conference on computer vision and
  pattern recognition workshops}, pages 136--144, 2017.

\bibitem{liu2018receptive}
Songtao Liu, Di Huang, et~al.
\newblock Receptive field block net for accurate and fast object detection.
\newblock In {\em Proceedings of the European Conference on Computer Vision
  (ECCV)}, pages 385--400, 2018.

\bibitem{schulter2015fast}
Samuel Schulter, Christian Leistner, and Horst Bischof.
\newblock Fast and accurate image upscaling with super-resolution forests.
\newblock In {\em Proceedings of the IEEE Conference on Computer Vision and
  Pattern Recognition}, pages 3791--3799, 2015.

\bibitem{shi2016real}
Wenzhe Shi, Jose Caballero, Ferenc Husz{\'a}r, Johannes Totz, Andrew~P Aitken,
  Rob Bishop, Daniel Rueckert, and Zehan Wang.
\newblock Real-time single image and video super-resolution using an efficient
  sub-pixel convolutional neural network.
\newblock In {\em Proceedings of the IEEE conference on computer vision and
  pattern recognition}, pages 1874--1883, 2016.

\bibitem{sun2008image}
Jian Sun, Zongben Xu, and Heung-Yeung Shum.
\newblock Image super-resolution using gradient profile prior.
\newblock In {\em 2008 IEEE Conference on Computer Vision and Pattern
  Recognition}, pages 1--8. IEEE, 2008.

\bibitem{szegedy2015going}
Christian Szegedy, Wei Liu, Yangqing Jia, Pierre Sermanet, Scott Reed, Dragomir
  Anguelov, Dumitru Erhan, Vincent Vanhoucke, and Andrew Rabinovich.
\newblock Going deeper with convolutions.
\newblock In {\em Proceedings of the IEEE conference on computer vision and
  pattern recognition}, pages 1--9, 2015.

\bibitem{timofte2017ntire}
Radu Timofte, Eirikur Agustsson, Luc Van~Gool, Ming-Hsuan Yang, and Lei Zhang.
\newblock Ntire 2017 challenge on single image super-resolution: Methods and
  results.
\newblock In {\em Proceedings of the IEEE conference on computer vision and
  pattern recognition workshops}, pages 114--125, 2017.

\bibitem{timofte2013anchored}
Radu Timofte, Vincent De~Smet, and Luc Van~Gool.
\newblock Anchored neighborhood regression for fast example-based
  super-resolution.
\newblock In {\em Proceedings of the IEEE international conference on computer
  vision}, pages 1920--1927, 2013.

\bibitem{wang2018recovering}
Xintao Wang, Ke Yu, Chao Dong, and Chen Change~Loy.
\newblock Recovering realistic texture in image super-resolution by deep
  spatial feature transform.
\newblock In {\em Proceedings of the IEEE conference on computer vision and
  pattern recognition}, pages 606--615, 2018.

\bibitem{wang2018esrgan}
Xintao Wang, Ke Yu, Shixiang Wu, Jinjin Gu, Yihao Liu, Chao Dong, Yu Qiao, and
  Chen Change~Loy.
\newblock Esrgan: Enhanced super-resolution generative adversarial networks.
\newblock In {\em Proceedings of the European Conference on Computer Vision
  (ECCV)}, pages 0--0, 2018.

\bibitem{yan2015single}
Qing Yan, Yi Xu, Xiaokang Yang, and Truong~Q Nguyen.
\newblock Single image superresolution based on gradient profile sharpness.
\newblock {\em IEEE Transactions on Image Processing}, 24(10):3187--3202, 2015.

\bibitem{yang2010image}
Jianchao Yang, John Wright, Thomas~S Huang, and Yi Ma.
\newblock Image super-resolution via sparse representation.
\newblock {\em IEEE transactions on image processing}, 19(11):2861--2873, 2010.

\bibitem{yang2019deep}
Wenming Yang, Xuechen Zhang, Yapeng Tian, Wei Wang, Jing-Hao Xue, and Qingmin
  Liao.
\newblock Deep learning for single image super-resolution: A brief review.
\newblock {\em IEEE Transactions on Multimedia}, 21(12):3106--3121, 2019.

\bibitem{zeyde2010single}
Roman Zeyde, Michael Elad, and Matan Protter.
\newblock On single image scale-up using sparse-representations.
\newblock In {\em International conference on curves and surfaces}, pages
  711--730. Springer, 2010.

\bibitem{zhang2020ntire}
Kai Zhang, Shuhang Gu, Radu Timofte, et~al.
\newblock Ntire 2020 challenge on perceptual extreme super-resolution: Methods
  and results.
\newblock In {\em IEEE Conference on Computer Vision and Pattern Recognition
  Workshops}, 2020.

\bibitem{zhang2018image}
Yulun Zhang, Kunpeng Li, Kai Li, Lichen Wang, Bineng Zhong, and Yun Fu.
\newblock Image super-resolution using very deep residual channel attention
  networks.
\newblock In {\em Proceedings of the European Conference on Computer Vision
  (ECCV)}, pages 286--301, 2018.

\end{thebibliography}
}

\end{document}